\documentclass[useAMS,usenatbib]{mn2e}
 \usepackage[dvips]{graphicx}
 \usepackage[english]{babel}

\title[Clues on the origin of galactic spin]
{Clues on the origin of galactic angular momentum from looking at galaxy pairs}
\author[B. Cervantes-Sodi, X. Hernandez and Changbom Park]{B. Cervantes-Sodi$^{1,2}$\thanks{E-mail: bcsodi@kasi.re.kr}, X. Hernandez$^{1,3}$\thanks{E-mail: xavier@astroscu.unam.mx} and 
Changbom Park$^{4}$\thanks{E-mail: cbp@kias.re.kr}\\
$^{1}$Instituto de Astronom\'\i a,
Universidad Nacional Aut\'onoma de M\'exico
A. P. 70--264,  M\'exico 04510 D.F., M\'exico \\
$^{2}$Korea Astronomy and Space Science Institute, 61-1 Hwaam-dong, Yuseong-gu, Daejeon 305-348, Korea \\
$^{3}$GEPI, Observatoire de Paris, Meudon Cedex, France\\
$^{4}$Korea Institute for Advanced Study, Dongdaemun-gu, Seoul 130-722, Korea\\
}
\begin{document}

\date{In original form 2009 August 18}

\pagerange{\pageref{firstpage}--\pageref{lastpage}} \pubyear{2009}

\maketitle

\label{firstpage}

\begin{abstract}

We search for correlations between the spin in pairs of spiral galaxies, to study if the angular 
momentum gain for each galaxy was
the result of tidal torques imprint by the same tidal field. To perform our study we made use of a sample of
galaxy pairs identified using the Sloan Digital Sky Survey. We find a weak, but statistically significant
correlation between the spin magnitude of neighbouring galaxies, but no clear alignment between their orientation.
We show that events such as interactions with close neighbours 
play an important role in the value of the spin for the final configuration, as we find these interactions
tend to reduce the value of the $\lambda$ spin parameter of late-type galaxies considerably,
with dependence on the morphology of the neighbour.
This implies that the original tidal field for each pair could have been similar, but 
the redistribution of angular
momentum at later stages of evolution is important. 
\end{abstract}


\section{Introduction}

A fundamental issue in the study of galaxy formation and evolution is the internal angular
momentum, from its acquisition to identifying the role it plays in determining the final
structure of galactic systems.

To explain the origin of the internal angular momentum in galaxies, Hoyle (1949) proposed the tidal
field theory, later studied by Peebles (1969) and White (1984).
In this theory, galaxies acquire their angular momentum by
gravitational torquing exerted, in the early Universe, by neighbour protogalaxies.
The tidal fields from the large-scale structure surrounding the protogalaxy exert a torquing moment 
increasing the angular momentum of the
system, which grows linearly in time prior to the gravitational collapse of the structure as  
protogalaxies have large linear sizes, and thus, the
surrounding material is able to imprint angular momentum efficiently. At 
later stages, after the protogalaxy decouples from the expanding background and turns around, the 
growth of angular momentum is reduced to a second order effect,
because the collapse dramatically reduces the lever arms (Schaefer 2009).

Several analytical and numerical studies have been performed to study this theory. The first concern
was to quantify the amount of angular momentum imprinted on the galaxies (Peebles 1969; Doroshkevich 1970;
 White 1984; Barnes \& Efstathiou 1987). It is usual
to express the total angular momentum of the galaxies through the dimensionless angular momentum parameter
 $\lambda$, which in standard cosmological models peaks at values
around $\lambda_{0} \approx 0.04$ and shows a log-normal distribution with a spread in the logarithm of 
 $0.48<\sigma_{\lambda}<0.66$ (for a
compilation of theoretical estimates on $\lambda_{0}$ and $\sigma_{\lambda}$ see Shaw et al. 2006).
In consistence with what was found in the indirect studies of Syer, Mao \& Mo (1999), the estimates 
of the distribution of $\lambda$ for cosmological dark matter haloes from analytical calculations and 
N-body simulations, have recently been shown to be in agreement with inferences of this quantity 
from observed galactic structural parameters for large volume limited samples, by Hernandez et al. (2007) 
and Berta et al. (2008).

Since the introduction of the $\lambda$ parameter, and with the boom of high resolution N-body 
simulations and detailed semi-analytical models,
a well stocked collection of studies has emerged, some looking at the evolution of the spin parameter
(Warren et al. 1992; Bullock et al. 2001; Vitvitska et al. 2002; Pierani, Mohayaee \& de Freitas 
Pacheco 2004, Davis \& Natarajan 2009), dependencies with
environment (Avila-Reese et al. 2005; Maccio et al. 2007; Bett et al. 2007) and correlations with 
virial mass (Barnes \& Efstathiou 1987;
Cole \& Lacey 1996; Tonini et al. 2006; Bett et al. 2007, Cervantes-Sodi et al. 2008), and others concerning the direct 
influence of this parameter on the internal
structure of the galaxies (Fall \& Efstathiou 1980, Mo, Mao \& White 1998, van den Bosch 1998, 
Jimenez et al. 1998, Prantzos \& Boissier 2000, Boissier et al. 2001, Hernandez \& Cervantes-Sodi 2006, 
Berta et al. 2008, Cervantes-Sodi \& Hernandez 2009). Most of this works deal only with
numerical simulations, lacking observational counterparts.
This, as the angular momentum of the dominant dark haloes
can not be directly measured in real galaxies, although some
recent efforts have begun to address the issue of obtaining
large observational samples as counterparts (e.g. 
Berta et al. 2008; Cervantes-Sodi et al. 2008).
The orientation of the angular momentum, however, is easier to obtain, if one assumes that
the shape of the galaxies is largely coupled to the overall angular momenta of the haloes.

The interest concerning possible alignments between galaxies due to the mechanism of
acquisition of angular momentum, has motivated several observational studies.
Gott \& Thuan (1978), exploring the angular momentum in the local group, pointed out that
if a pair of galaxies (like the Galaxy and M31) is formed in isolation,
and the collapse phase of formation was short compared with other dynamical time scales,
the spin vectors of the galaxies should be perpendicular to the separation vector between
the galaxies and parallel between them. This, for the particular case of the local group,
holds rather well. Sharp, Lin \& White (1979) extended the study to a sample of nearly 100 pairs and 
reported a complete lack of correlation between the spins. Oosterloo (1993) reached the
same conclusion, contrary to the result obtainned by Helou (1984), who found an asymmetric
correlation; they report that the spin vectors avoid being parallel, in favour of the anti-parallel configuration.
More recently, Pesta\~na \& Cabrera (2004) studied the relative alignment of spin axes for
observed pairs of spiral galaxies, and found some significance against the null hypothesis of
random orientations of binary spiral galaxies. They conclude that complex and repeated interactions
probably occur in binary spiral galaxies, so that spins that are parallel, antiparallel or 
nearly orthogonal often occur. Slosar et al. (2009) detected a correlation 
in the spin directions of pairs of spiral galaxies, with significant correlation at 
small separations ($<0.5Mpc/h$).

With the aim
of elucidating the process of galaxy assembly,
some studies concerning alignments of galaxies in clusters have also been performed. It is believed
that galaxies and clusters are formed through a process of hierarchical accretion. In
that case, primordial alignments could quickly be erased by dynamical interactions (Coutts 1996;
Plionis et al. 2003), and the current alignments would be produced by recent accretions
episodes (Faltenbacher et al. 2005). The positive detection of alignments seems to depend
on several characteristics of the clusters, such as the presence of substructure (Plionis
\& Basilakos 2002), its accretion history (Faltenbacher et al. 2005), the morphology of the galaxies
(Faltenbacher et al. 2007; Torlina, De Propris \& West 2007; Wang et al. 2009)
and the morphology of the cluster itself (Aryal \& Saurer 2006; Aryal, Paudel \& Saurer 2007), which
reflects not only the initial conditions, but the dynamical interactions taking place in such
complex systems.

Using numerical simulations, Barnes \& Efstathiou (1987) fixed their attention on this
issue, and found no consistent indication of coherence in the spins of adjacent objects.
The result of Bailin \& Steinmetz (2005) is similar, they could not find a clear tendency for the 
principal axes of neighbouring haloes
to point in a preferred direction, and Porciani, Dekel \& Hoffman (2002) argue that spatial 
correlation of spins on scales larger than
a few Mpc, induced by primordial tidal torques, are strongly affected by non-linear effects.

In this paper we extend the study of angular momentum correlations between neighbouring galaxies, 
using the $\lambda$ parameter
to account for the magnitude of the spin of observed galaxies.
This adds an extra variable to the analysis of angular momentum in observed galaxy pairs, which had up to now
been limited to considering only the orientation of the galaxies. A more quantitative description
of the problem is hence possible, allowing for a more detailed comparison to the outcome of
cosmological simulations, and yielding interesting constraints on the origin of spiral galaxy pairs.
As a first test we begin with the simplest case, that of galaxy pairs. This is a regime which makes our work complementary to existing studies in clusters e.g. Faltenbacher et al. (2005) or Wang et al. (2009).
Extensive studies on the effects of the large-scale environment and interactions with neighbouring
galaxies on various other galaxy properties are presented by Park, Gott \& Choi  (2008, hereafter PGC)
and Park \& Choi (2009).

The plan of this paper is as follows. In Section 2, we review
 the derivation of the $\lambda$
spin parameter for inferred haloes of spiral galaxies as developed in Hernandez \& Cervantes-Sodi (2006),
in Section 3 we present the SDSS sample of galaxy pairs used in the present work. Section 4  
presents the results from the analysis, and finally, in Section 5, we summarize our general conclusions.

\section{Estimation of the spin from observable parameters}

To quantify the magnitude of the angular momentum, we employ the $\lambda$ spin parameter as 
introduced by Peebles (1969); 

\begin{equation}
\label{Lamdef}
\lambda = \frac{L \mid E \mid^{1/2}}{G M^{5/2}}
\end{equation}

where $E$, $M$ and $L$ are the total energy, mass and angular momentum of the configuration, 
respectively. In 
Hernandez \& Cervantes-Sodi (2006) we derived a simple estimate of total $\lambda$ for dark halos
hosting disc galaxies 
in terms of observational parameters, and 
showed some clear correlations between this parameter and structural parameters, such as the disc to
 bulge ratio,
the scale height and the colour. Here we recall briefly the main ingredients of this model.
 The model 
considers only two components for galaxies, a disc for the baryonic component with an exponential surface mass
 density $\Sigma(r)$;

\begin{equation}
\label{Expprof}
\Sigma(r)=\Sigma_{0} e^{-r/R_{d}},
\end{equation} 

where $r$ is a radial coordinate and $\Sigma_{0}$ and $R_{d}$ are two constants which are allowed 
to vary from
galaxy to galaxy, and a dark matter halo having an isothermal density profile $\rho(r)$,
 responsible for establishing
a rigorously flat rotation curve $V_{d}$ throughout the entire galaxy;

\begin{equation}
\label{RhoHalo}
\rho(r)={{1}\over{4 \pi G}}  \left( {{V_{d}}\over{r}} \right)^{2}. 
\end{equation}

We assume: (1) that originally the baryonic and dark matter components of the protogalaxy
were well mixed, aquiring the same specific angular momentum, as often asumed in
galactic formation models (e.g. Fall \& Efstathiou 
(1980); Mo et al. (1998)), and recently confirmed by several authors addressing this particular
issue (Zavala et al. (2008); Croft et al. (2008); Piontek \& Steinmetz (2009)); 
(2) the total energy is dominated by that of the halo which is a virialized gravitational structure; 
(3) the disc 
mass is a constant fraction of the halo mass $F=M_{d}/M_{H}$. These assumptions allow us to express 
$\lambda$ as

\begin{equation}
\label{Lamhalo}
\lambda=\frac{2^{1/2} V_{d}^{2} R_{d}}{G M_{H}}.
\end{equation}

Finally, we introduce a baryonic disc Tully-Fisher (TF) (e.g  Gurovich et al. 2004) relation: 
$M_{d}=C_{TF} V_{d}^{3.5}$, and taking the Milky Way as a 
representative example, we evaluate $F$ and $C_{TF}$ to obtain

\begin{equation}
\label{LamObs}
\lambda=21.8 \frac{R_{d}/kpc}{(V_{d}/km s^{-1})^{3/2}}.
\end{equation}

In Cervantes-Sodi et al. (2008) we tested the accuracy of our estimate of $\lambda$, comparing
 the value obtained
using equation ~\ref{LamObs} to the actual value of $\lambda$ from numerically simulated 
galaxies from six distinct
groups, where the actual value of $\lambda$ is known and where we had also estimated it through 
equation ~\ref{LamObs}, as
baryonic disc scale lengths and disc rotation velocities for the resulting simulated galaxies
were given. The test showed an unbiased and tight one-to-one correlation, with very small 
dispersion leading to errors  $< 30\%$. Application to a large volume limited sample of
over 11,000 galaxies from the SDSS yielded 
a log-normal distribution of $\lambda$ for the total sample, interestingly in consistence with
results from cosmological n-body simulations, presented in Hernandez et al. (2007). 
The above results were then reproduced by Berta et al. (2008), using a larger sample of 50,000 galaxies
also from the SDSS, and a refined version of the $\lambda$ estimate given here, for spiral galaxies.

\section{The SDSS sample}

The sample of galaxy pairs used in this work comes form a study by Park, Gott \& Choi (2008, hereafter
PGC), using
data from the SDSS.
It is a volume-limited sample of galaxies with absolute magnitude $M_r<-19.5 + 0.5{\rm log} h$ in the 
redshift interval $0.001 < z < 0.5$.  Since most theoretical studies concerning spin distributions 
present their 
results at $z=0$, we limited the sample to low redshifts.

The nearest neighbour for a given galaxy is found requiring the following conditions; (1) the neighbour 
galaxy can not
be fainter than the target galaxy by more than $\Delta M_{r}$, (2) it must have the smallest projected 
separation across the line of sight from the target galaxy and (3), it must present a radial velocity difference less 
than $V_{max}$. We choose $\Delta M_{r}=0.5$ to include only influential neighbours and not merely satellite galaxies. 
In PGC it was shown that the selection of target galaxies having neighbors fainter
by more than 0.5 mag produces similar results but drastically  reduces the number 
of target galaxies as their absolute magnitude cut becomes brighter, which reduces
the statistical significance of the results. 

To determine $V_{max}$, PGC searched for all neighbours with a velocity difference of $<1000 km s^{-1}$ 
with respect to each target galaxy and with a magnitude not fainter by more than $\Delta M_{r}$.
When looking at the rms velocity difference of the neighbours as a function of the projected
separation, this remains constant out to $50 h^{-1}$ kpc, at 255 and 169 $km s^{-1}$ for early and
late type target galaxies respectively. In this way, we adopt $V_{max}=$ 600 and 400 $km s^{-1}$ for
early and late type galaxies, limits that correspond to about 2.3 times the rms values.

Given that the limiting magnitude of the
sample is $M_{r}=-19.0+5\log{h}$, we study only those target galaxies brighter than $M_{r}=-19.5+5\log{h}$, in 
order to avoid losing neighbours, we fixed
our attention on pairs of relatively bright galaxies.
For more details of the sample see Park, et al. (2008). For the same sample, Choi et al. (2007)
have determined the exponential disc scale, absolute magnitude, velocity
dispersion, de Vacouleurs radius and eccentricity for each galaxy, assuming a $\Lambda CDM$ 
universe with  $\Omega_{M}=0.27$, $\Omega_{\Lambda}=0.73$ and $h=0.71$. 

In order to discriminate between elliptical and disc galaxies, we used the prescription of
Park \& Choi (2005) in which early (ellipticals and lenticulars) and late (spirals) types are
segregated in a $u - r$ colour versus $g - i$ colour gradient space and in the concentration index space. 
They tested extensively the selection criteria through direct comparison of visually assigned types
for a large sample of several thousand galaxies. The specific selection criteria can be found in Park \& Choi (2005),
but essentially select as early types, galaxies with red colours, minimal colour gradients and high
concentration indices. Throughout the paper we will consider ellipticals and lenticulars as early type 
galaxies, and spirals as late type ones.

In the next section we will examine the correlation between the angular momenta of galaxies in pairs, 
taking into account angular momentum
magnitude through the $\lambda$ spin parameter, and direction through position angle. Later we
will investigate if the angular momentum of a given galaxy is modified by the presence of a neighbour. To 
conduct these inquiries, we obtained four different subsamples according to the following criteria.

Our sample A, used to search for a correlation in the $\lambda$ value of spiral galaxy pairs, 
contains 347 pairs of late type late - type galaxies, with $0.001 < z < 0.5$, limiting magnitude of 
$M_{r}=-19.5+5{\rm log} h$, and $R_{d}$ measured by Choi et al. (2007), and rotational 
velocities infered from the absolute
magnitude introducing a TF relation (Pizagno et al. 2007), which is all the information needed to calculate 
$\lambda$ from equation ~\ref{LamObs}. To avoid the problem of internal 
absorption in edge-on galaxies (Unterborn \& Ryden 2008; Cho \& Park 2009), and consequently underestimating rotational 
velocities, we limit the sample to spiral galaxies having axis ratios $b/a > 0.6$.

To study the correlation of the angular momentum direction between members of a spiral pair, we constructed
sample B, composed of 218 pairs, with the same redshift range and limiting absolute magnitude as sample A. 
Supposing the position angle of observed galaxies to be
perpendicular to their angular momentum vector, we have an indicator of the angular momentum orientation,
albeit with a degeneracy between parallel and anti parallel spins.
Given the difficulty of measuring the position angle for face-on late-type
galaxies, we used pairs of spiral galaxies with $b/a<0.7$. This is the reason why we could not use the same
sample A to search for a correlation in the direction of the angular momentum.

In both samples, A and B, we required that the nearest neighbour of a given neighbour galaxy in a galaxy pair, 
should be
the target galaxy itself, in this way we search for spin correlations for pairs involving only two galaxies, the simplest scenario.
To test the response of the spin to the presence of a companion, we relaxed both this condition and the condition of the
neighbour galaxy being a late type galaxy, which increases the number of galaxy pairs to 3624 for sample C,
which will be used to measure the response of the value of $\lambda$ to the distance of the nearest neighbour,
and to 2037 for sample D with
the position angle well determined for every galaxy involved, to determine the influence of a neighbour on alignment.

When studying the influence of a nearby galaxy on the value of the spin, the distance between galaxies is
normalized to the virial radius of the neighbour.
We define the virial radius of a galaxy as the projected radius where the mean mass
density within the sphere of radius $r_{vir}$ is 200 times the critical density or
740 times the mean density of the universe.

\begin{equation}
\label{Rvir}
r_{vir}^{3}= \left( \frac{3 }{4 \pi}\right) \left(  \frac{ 200 \gamma L }{ \rho_{c} }\right)   ,
\end{equation}

with the relative mass-to-light ratios for early type galaxies (ellipticals and lenticulars) twice 
the same ratio for the late types (spirals), $\gamma (early) = 2\gamma (late)$, following Choi et al. (2007), 
who report that the central velocity dispersion of early type galaxies
brighter than $M_{r}=-19.5$, is about $\sqrt{2}$ times that of late types. Since we adopt $\Omega_{M}=0.27$,
we have $200 \rho_{c}= 200 \overline{\rho}/\Omega_{M}=740\overline{\rho}$, which is almost equal to the virialized
density $\rho_{vir}= 18 \pi ^{2} / \Omega_{m}(H_{0}t_{0})^{2}\overline{\rho}=766\overline{\rho}$, in the
case of a $\Lambda$CDM universe (Gott \& Rees 1975). Finally, we introduced the mean value for the density of
the Universe, $\overline{\rho} = (0.0223 \pm 0.0005)(\gamma L)_{-20}(h^{-1}Mpc)^{-3}$, where $(\gamma L)_{-20}$
is the mass of a late type galaxy with $M_{r}=-20$ (Park et al. 2008).
In this way, the virial radii
of galaxies with $M_{r}=-19.5$, -20.0 and -20.5 are 260, 300 and 350$h^{-1}$ kpc for early types and
210, 240 and 280$h^{-1}$ kpc for late types, respectively.

Having limited ourselves to such relatively massive systems, which exclude any dwarf galaxies, also 
gives confidence on the validity of the constant baryon fraction hypothesis used in the estimates 
of $\lambda$. The above, as small systems where substantial mass loss due to the feedback effects of star 
formation probably applies, are not included.

\section{Results}
The stringent selection criteria described above ensure that our samples contain only galaxies where 
our estimates of the spin are most reliable, although the samples are reduced to a small fraction of 
the total SDSS field, they remain statistically significant.

The results of our $\lambda$ estimates and position angle studies appear in this section.
Section 4.1 tests the angular momentum acquisition mechanism through
an exploration of the correlations in magnitude and orientation of the angular momentum for close pairs
of spiral galaxies. In section 4.2 we investigate the effect on
spin magnitude and orientation, of the interaction of a spiral galaxy with another galaxy, as a function of the 
separation between them, regardless of the morphology of the companion.

\subsection{Spin correlations}

\begin{figure}
\label{LMplot}
\centering
\begin{tabular}{c}
\includegraphics[width=0.475\textwidth]{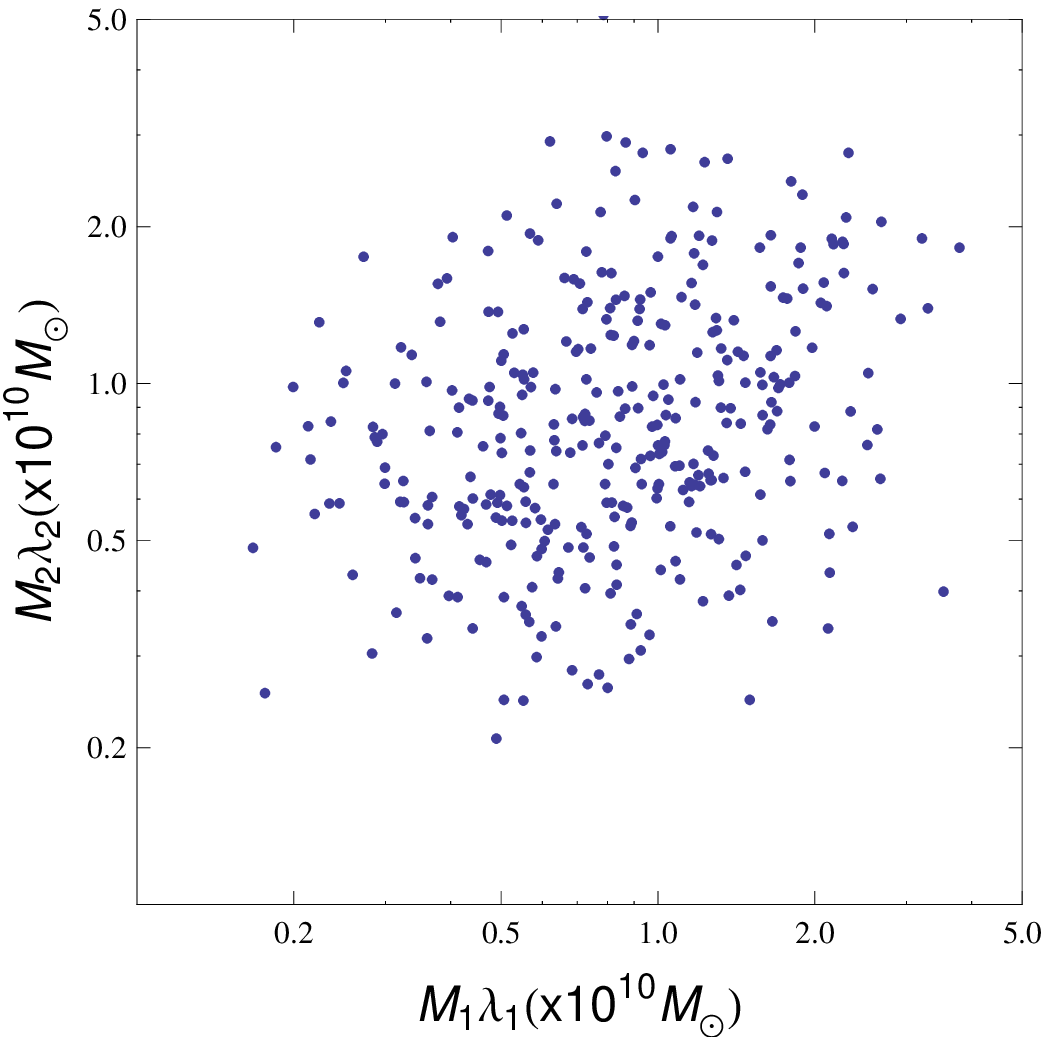}  \\
\includegraphics[width=0.475\textwidth]{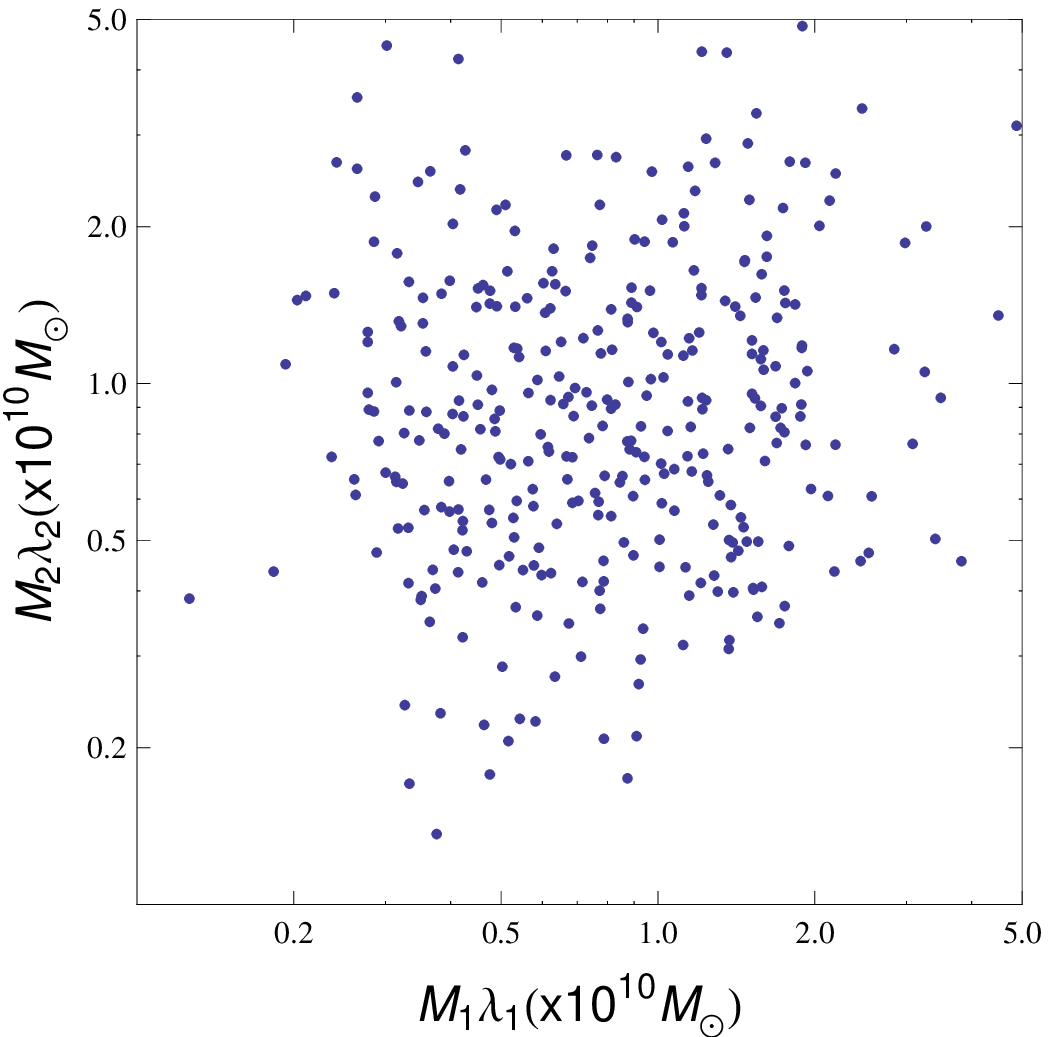}
\end{tabular}
\caption[ ]{$\lambda_{1} M_{1}$ of target galaxies against  $\lambda_{2} M_{2}$ of nearest neighbour 
galaxies, in units of solar masses, for 347 pairs. \textit{Top panel:} Pairs of sample A with a $r^{2}=0.247$. 
\textit{Bottom panel:}for comparison, result from a Monte Carlo sample with no inherent correlation 
beyond what is imposed by the magnitude selection criteria in the observed sample, with $r^{2}=0.045$.}
\end{figure}

It is reasonable to think that, given the proximity of the galaxies in each pair, both galaxies
are immersed in the same tidal field. The amount of angular momentum gained by each galaxy 
should be similar, proportional to the tidal torques
exerted by the surrounding material and inversely proportional to the mass of each galaxy. 
This assumes that the two components of the present pair have been in close association, at least 
since the phase of angular momentum acquisition. We explore this hypothesis with 
the selected sample of pairs of spiral galaxies, calculating spin parameters using equation 
~\ref{LamObs}, and the total mass from the
baryonic Tully-Fisher relation introduced to obtain equation ~\ref{LamObs}.

If both member galaxies of a pair were exposed to the influence of the same tidal
field, their product $\lambda M$ should show a clear and tight 1 to 1 correlation.
In Fig. 1 top panel is shown the spin mass product of nearest neighbour galaxies $M_{2}\lambda_{2}$, as a 
function of the same product for target galaxies $M_{1}\lambda_{1}$, in units of solar masses 
for the 347 pairs of late type - late type galaxies extracted from our sample A.
In the scenario where both galaxies gained their
angular momentum through interactions with a constant tidal field, we should see a clear correlation, but the correlation 
between the products for neighbouring galaxies in this case is low, with a correlation index of $r^{2}=0.247$.

To quantify the low level correlation seen in the top panel of figure 1, we produced a large number 
of equivalent mock catalogues of spiral-spiral galaxy pairs through Monte Carlo simulations.
Each, with a total of 347 pairs, was obtained through sampling a fixed distribution
function for both of the numbers which describe the members of an observed pair. A random value
of $M$ and $\lambda$ is picked for each member of a sampled pair, until one obtains as many
mock pairs as present in our sample A.

Given the restriction in the absolute magnitude imposed when selecting the observed
pairs, the first condition to find the nearest neighbour, a constraint is imposed upon 
the difference in mass within the members of a given pair. The mass input for the test
was chosen observing this restriction, using a constant density distribution with upper and lower limits 
for the mass as imposed on the SDSS pairs sample. 
The values of $\lambda$ were taken from the log-normal
distribution obtained empirically in Hernandez et al. (2007) with parameters $\lambda_{0}=0.0394$ and 
$\sigma_{\lambda}=0.509$, for a larger sample of SDSS galaxies. A mock sample is shown in
Fig. 1 bottom panel, for comparison with the sample A of galaxy pairs from the SDSS,
showing a correlation index of $r^{2}=0.135$.
In this particular case, the correlation index is lower for the mock sample,
perhaps the result of an statistical fluctuation. To quantify this effect we measured the 
correlation index for one thousand mock catalogues of pairs, all with 347 pairs,
and found a mean value of $\langle r^{2} \rangle = 0.129$ and, given that the 
collection of synthetic correlation
indexes is well described by a normal distribution with a spread of 0.063,
it places the correlation between the SDSS pairs about $1.9\sigma$ above the expected for null
intrinsic correlation between $\lambda_{1}M_{1}$ and $\lambda_{2}M_{2}$. Even though the correlation
for the sample of SDSS galaxy pairs seems weak and the correlation index is low, the correlation
is significantly stronger than that obtained for null intrinsic correlation mock samples. We must take into
account that the correlation extends over 1.5 magnitudes in the $M\lambda$ product, and that this
relation is dominated by galaxy pairs with large separations.

\begin{table}
\caption{Correlation indexes for Sample A galaxies and mock catalogues, showing for the
latter $\langle r^{2} \rangle \pm \sigma$ values.}
\begin{tabular}{|c|cc|}

	\hline

    $r^{2}$ & Sample A    &   mock catalogues    \\
	\hline
	$M_{2}\lambda_{2}$ vs $M_{1}\lambda_{1}$ & $0.247$  & $0.129 \pm 0.063$ \\
	$\lambda_{2}$ vs $\lambda_{1}$ & $0.202$  & $0.0056 \pm 0.0604$ \\
	\hline
\end{tabular}
\end{table}

Due to the restriction in magnitude imposed when defining the sample of observed pairs, 
the difference in mass between each member of a pair is small, and allows us to
search for a direct correlation between $\lambda_{1}$ and $\lambda_{2}$. The result, using the same sample A,
shows no difference with the previous test, with a $r^{2}=0.202$, a weaker value but still larger than what is 
obtained with the results of a Monte Carlo test 
with no intrinsic correlation at all; with $\langle r^{2} \rangle = 0.0056$ and $\sigma=0.0604$.  It appears likely that 
the correlations shown in Table 1 have their origin in a common tidal field having being at least partly 
responsible for imprinting the angular momentum to both members of the pairs studied. However, the fact that 
this correlations are weak, points to other angular momentum acquisition and modification mechanisms having 
being at work, e.g. mergers.

So far we have centred our attention on the magnitude of the spin parameter, but we have not yet examined
the direction of the angular momenta in the pairs. 

For the analysis of the angular momentum direction, we use sample B.
We measured the difference in the position angle for the galaxies of each pair, restricted to vary 
in the range $0<\Theta<90$, given our impossibility to discriminate between parallel and 
anti-parallel spins. Once obtaining $\Theta$ for each pair, following the method proposed by Yang et al.
(2006), we count the total number of pairs, $N\left( \Theta \right) $, for a number of bins in $\Theta$.
Next, we construct 100 random samples in which we randomized the orientation of the galaxies, respecting
the selection criteria of the real sample, and we computed the average number of pairs $\left\langle N_{R} \left( \Theta 
\right) \right\rangle $, as a function of $\Theta$. By construction, the random samples have the same
selection effects as the real sample of pairs, so any significant difference between $N\left( \Theta \right) $
and $N_{R} \left( \Theta \right) $ reflects a genuine alignment between galaxy pairs.

To quantify the strength of any possible alignment, we define the normalized pair count as

\begin{equation}
f_{pairs} \left( \Theta \right) = \frac{N\left( \Theta \right)}{\left\langle N_{R} \left( \Theta 
\right) \right\rangle }.
\end{equation}

For a complete absence of alignment we should obtain $f_{pairs} = 1$. To assess the significance of
the deviation of the normalized pair count form unity, we use $\sigma_{R}\left( \Theta \right) /
\left\langle N_{R} \left( \Theta \right) \right\rangle$, where $\sigma_{R}\left( \Theta \right) $ is the
standard deviation of $N_{R}$ obtained form the 100 random samples.

\begin{figure}
\centering
\includegraphics[width=0.475\textwidth]{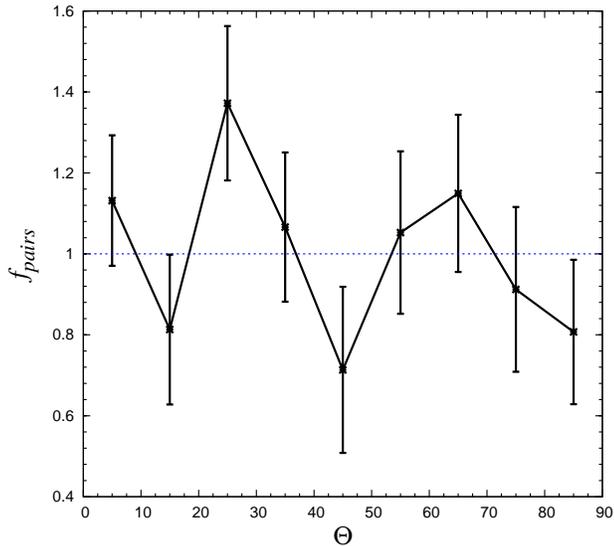}
\label{histograms}
\caption[ ]{ Normalized probability distribution of the difference in the position angle, $\Theta$, for the pairs
in sample B, dotted line representing $f_{pairs} = 1$, and the error bars give the value of
$\sigma_{R}\left( \Theta \right) /
\left\langle N_{R} \left( \Theta \right) \right\rangle$. }
\end{figure}

The result of applying this technique to our sample B is presented in Fig. 2, where we show $f_{pairs}$ for
the 255 pairs in the sample, and the value of $\sigma_{R}\left( \Theta \right) /
\left\langle N_{R} \left( \Theta \right) \right\rangle$ as error bars.
No intrinsic alignment is evident, as there are no significant deviations from
$f_{pairs} = 1$, represented by the dotted line, which is consistent with the result obtained with the
$\lambda$ magnitudes of the pairs; a very weak correlation between the spins of neighbouring spiral
galaxies, consistent with previous studies finding little correlation between 
the position angles of spiral-spiral pairs (Wang et al. 2009), which hence appear to have formed as a 
pair mostly after the phase of
angular momentum acquisition was over. In this particular case, the sample is dominated by pairs with
large separations; i. e. $d > R_{vir}$, as we will see in the next subsection.

The low correlation between the spin of neighbouring galaxies 
can be explained by the early acquisition of angular momentum through primordial torques exerted by the
surrounding tidal field on the pair (Barnes \& Efstathiou 1987; Navarro, Abadi \& Steinmetz 2004),
which induces an initial inherent correlation. In this sense, the correlations found (reported in table 1)
constitute an important direct empirical confirmation of the validity and importance of primordial tidal
torquing in contributing to galactic angular momentum acquisition. The effects of different subsequent 
mechanisms such as strong interactions with neighbour galaxies (Hernquist 1993), delayed mass 
aggregations events (Kitzbichler \& Saurer 2003; Pierani et al. 2004) and minor mergers (Gardner 2001; 
D'Onghia \& Navarro 2006) are to then blur the initial correlations, to the point where no evident 
alignment correlation remains, highlighting the effect of later processes. In the next subsection we 
will focus on the effect of interactions in the evolution of the spin.

\subsection{Interacting galaxies}

In this subsection we investigate if there is
a response to the interaction between neighbouring galaxies in the $\lambda$ spin
parameter, as a function of the separation distance between the members of the observed pairs. 
In this case we are only concerned
with the value of $\lambda$ for late type target galaxies, irrespective of the $\lambda$ characteristics
of the nearest neighbour. This allowed us to increment the number of systems studied, through
the inclusion of spiral target galaxies having early type closest neighbours, for the third of our tests.

Recently Park \& Choi (2009) reported a dependence of the absolute magnitude of the target galaxy on the
nearest neighbour separation, where the galaxy luminosity in the red decreases as the separation distance decrease.
Given that in the calculation of $\lambda$ we use $M_{r}$ to assign a $V_{d}$ value for equation
~\ref{LamObs}, we limited the sample to galaxies having magnitudes in the range
$-19.5 < M_{r} < -20.5$, to focus our study on the response of $\lambda$ not merely as a
consequence of the dependence found by Park \& Choi (2009) on magnitude, but as a directly consequence 
of the influence of a neighbour galaxy on the value of $\lambda$. 
Taking only a narrow range in $M_{r}$ limits the effect on the estimates of $\lambda$
of a correlation between $M_{r}$ and the nearest neighbour distance, to better isolate and asses 
the effects of the distance to the nearest neighbour, on
the spin of the target galaxy.

\begin{figure}
\begin{tabular}{c}
\includegraphics[width=84mm]{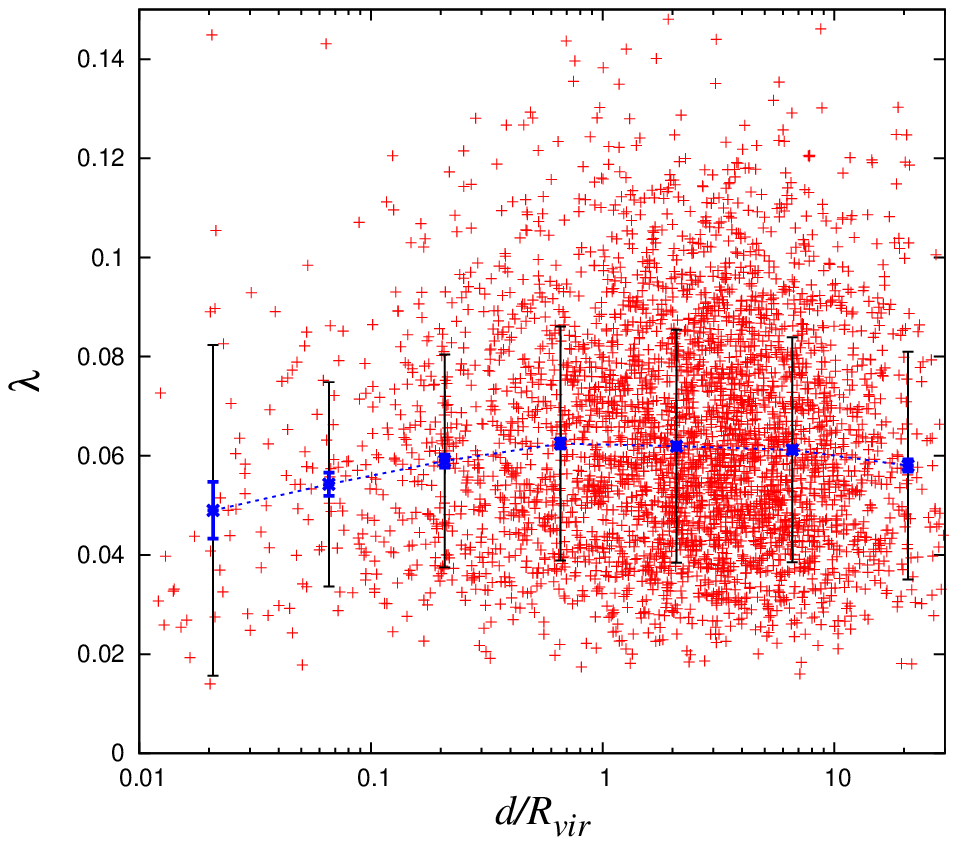}\\
\includegraphics[width=84mm]{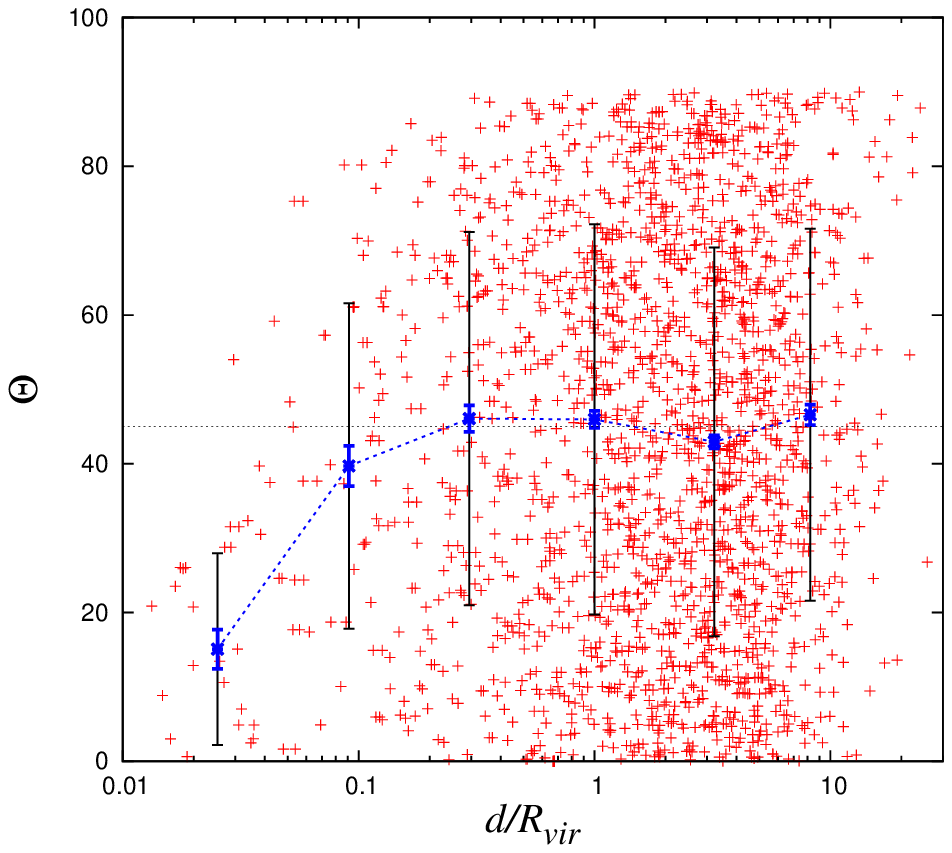}
\end{tabular}

\caption[ ]{ \textit{Top panel:} $\lambda$ value for the 3624 target late type from sample C as a
function of the separation distance
to their nearest neighbour, normalized by the virial radius of the neighbour galaxy, $R_{vir}$
\textit{bottom panel:} Difference in the position angle $\Theta$ for the 2037 late type pairs from sample D as a
function of their normalized separation distance $d/R_{vir}$}
\end{figure}

The top panel of figure 3 shows $\lambda$ values of the 3624 target late type
galaxies of sample C, as a function of the distance to their closest neighbours, normalized to the virial
radius of the neighbour galaxy, calculated using equation ~\ref{Rvir}. The sample is
divided into 7 bins, where the
median $\lambda$ values are shown with their dispersion presented as thin error
bars and the uncertainty represented by thick error bars; this
convention will be followed for the next figures.
For separations larger than the virial radius of the neighbour galaxy, the mean value of
$\lambda$ appears constant, but as soon as the distance becomes smaller than $R_{vir}$, 
the median value of $\lambda$ starts to decrease, a clear indication of the interaction. 
Galaxy harassment and various angular momentum loss mechanisms, such as dynamical friction (Hernquist 1993),
might begin to operate during the interaction,
even at large distances, as the galaxies first cross into their virial radii (PGC; Park
\& Choi 2009).

In going to the first bin, a substantial increase in the dispersion of the inferred values
of $\lambda$ is evident. However, for such small separations, systems within 0.02 of their virial radius,
a strong interaction is ongoing. This implies the systems are heavily disturbed and strongly out of
equilibrium; see e.g. the numerical simulations of galaxy collisions of Hernandez \& Lee (2004), where 
fluctuations in potential and kinetic energies of the total system are assessed as a function of the 
pair separation. This invalidates the assumptions going into the simple formula used to estimate $\lambda$,
making the values of this quantity in the first bin of figure 3, top panel, useful only as indications of
strong interactions at small distances, and not as indicative of total halo $\lambda$.

The complementary study using the difference in the position angle $\Theta$, to account
for the relative orientation
of the galaxies within a pair, is shown in Fig. 3 bottom panel using sample D, where
$\Theta$ is plotted against the separation distance between neighbouring galaxies, normalised
by the virial radius $R_{vir}$
of the neighbour galaxy. 
The sample is divided into 7 bins, where the
median $\Theta$ values are shown with their dispersion presented as thin black error
bars, as in the preceding plot.
If the orientations of the galaxies between neighbours were isotropic, we would expect a
median $\Theta$ value of
$45$, showed as a broken line in the plot. As can be seen, the median value
of $\Theta$ decreases as the separation distance
decreases below the virial radius of the neighbour galaxy, meaning that the
galaxies tend to be aligned (or
anti-aligned) as their neighbours get closer to them.

\begin{figure*}
\centering
\begin{tabular}{cc}
\includegraphics[width=.475\textwidth]{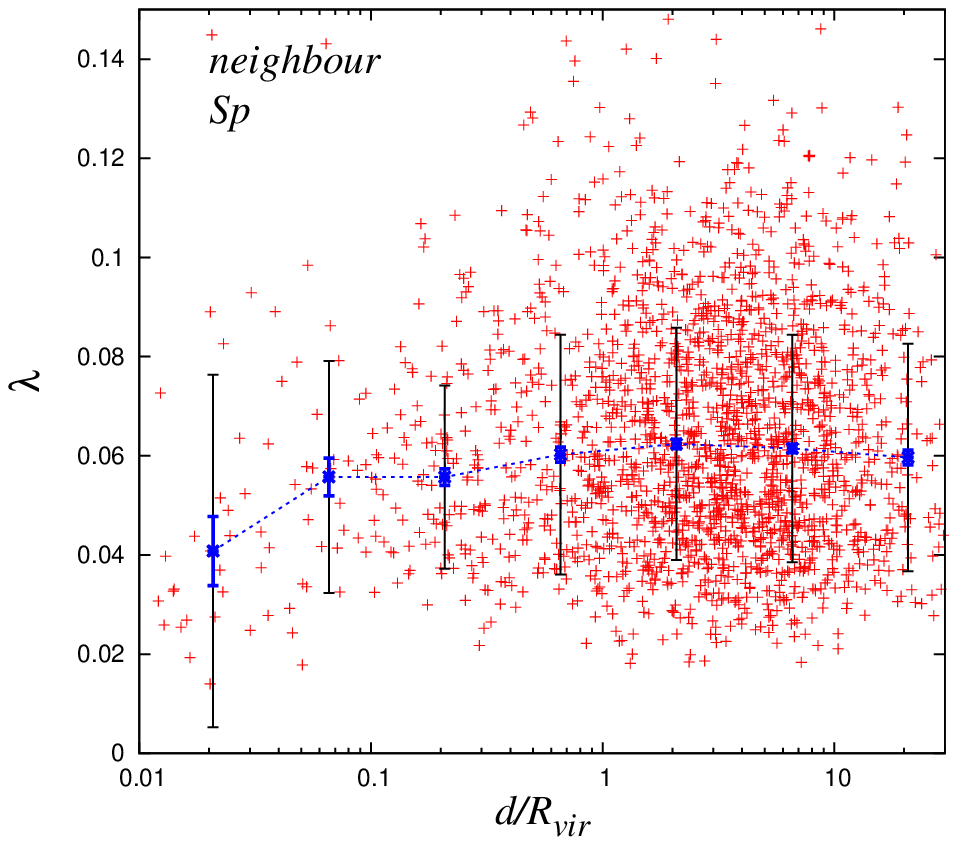} & \includegraphics[width=.475\textwidth]{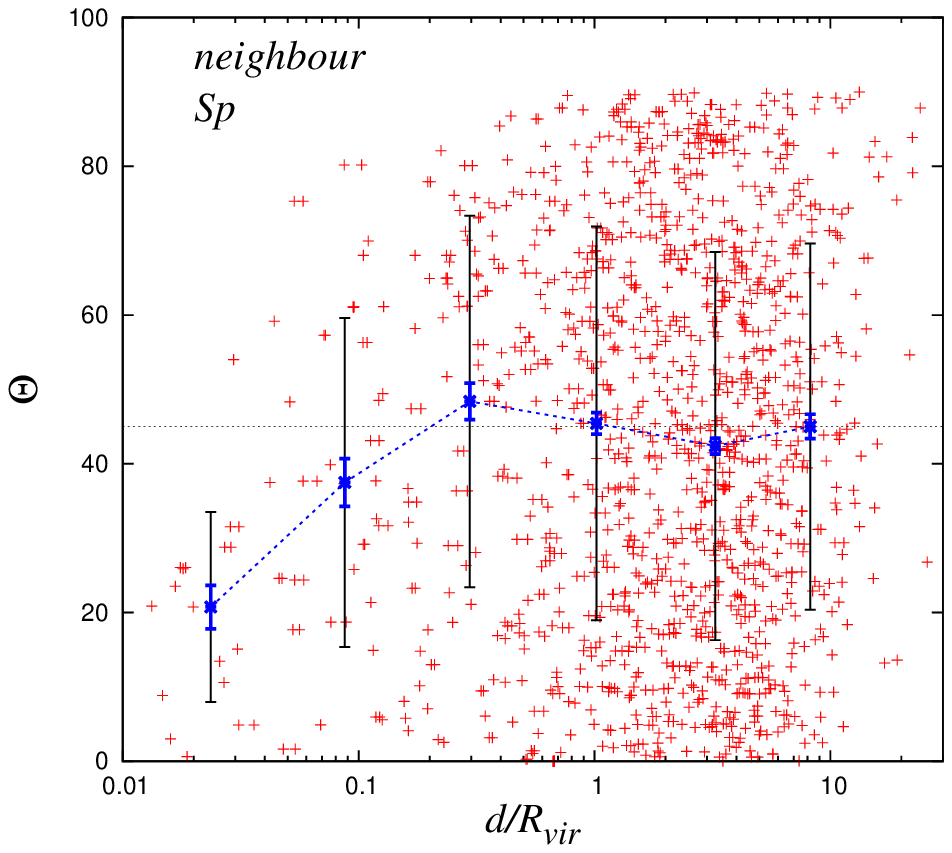}\\
\includegraphics[width=.475\textwidth]{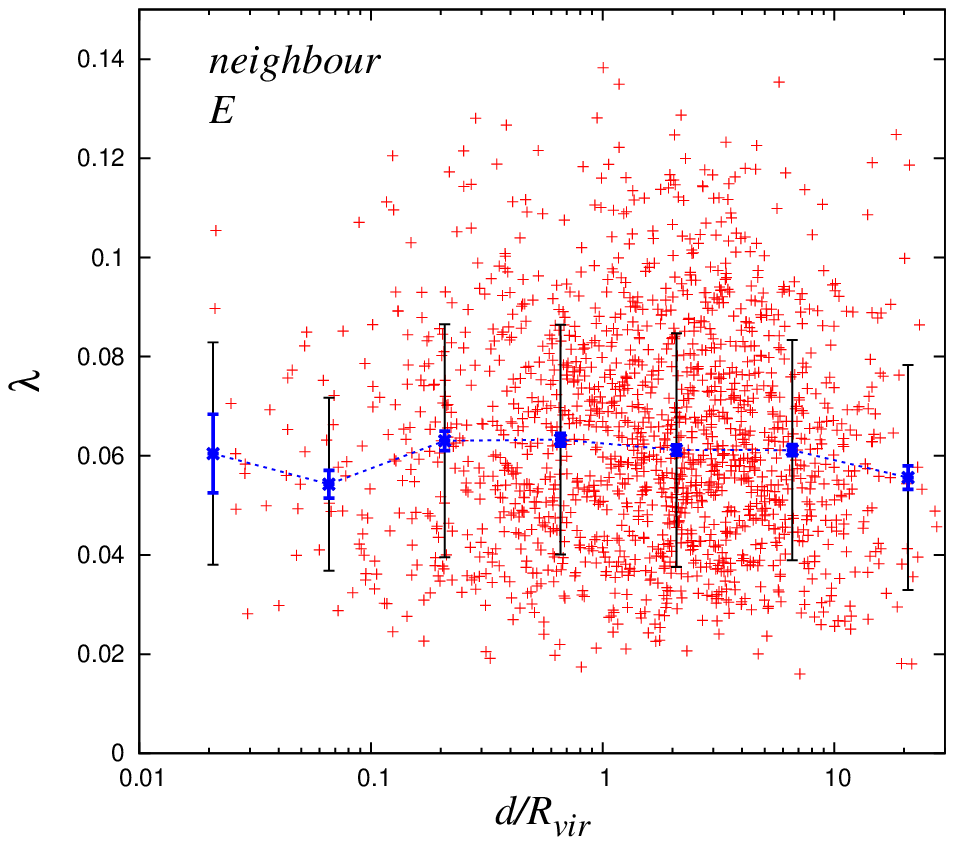}  & \includegraphics[width=.475\textwidth]{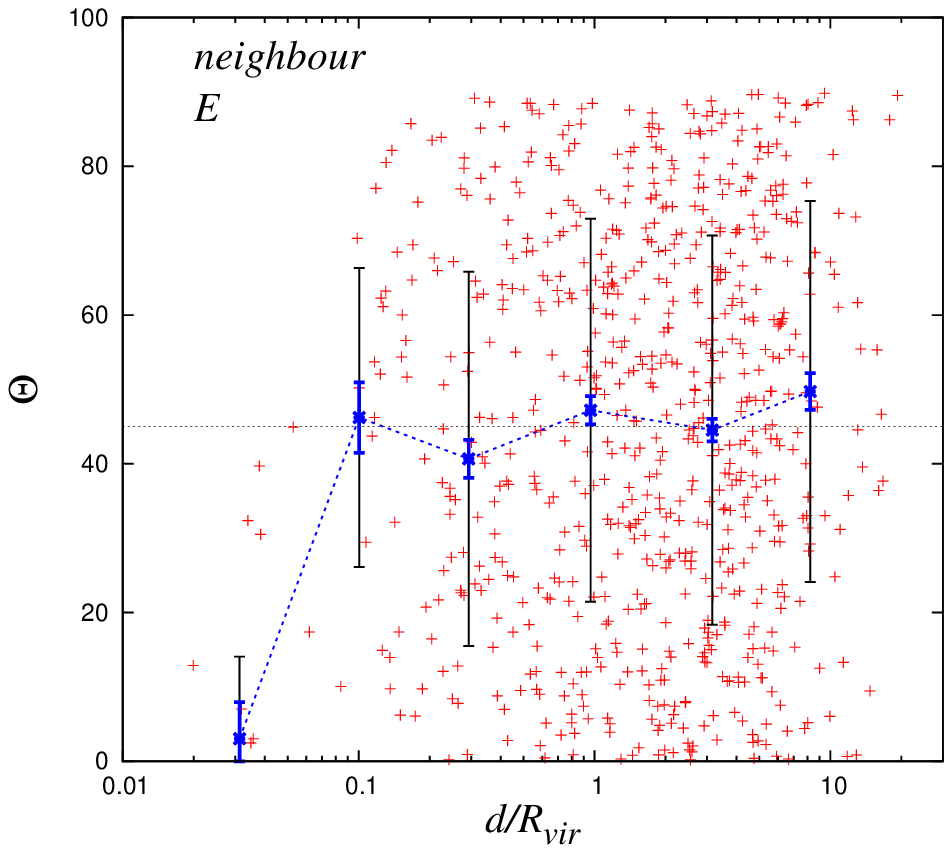}
\end{tabular}
\caption[ ]{\textbf{Left panels:} $\lambda$ value of target late type galaxies from sample C as a
function of the normalized separation distance
to their nearest neighbour, $d/R_{vir}$; \textit{top panel:}  nearest neighbour being a late type galaxy,
\textit{bottom panel:}  nearest neighbour being an early type galaxy.

\textbf{Right panels:} difference in the position angle for galaxies from sample D as a function
of the normalized separation distance to the nearest neighbour; \textit{top panel:}  nearest neighbour being a late type galaxy,
\textit{bottom panel:}  nearest neighbour being an early type galaxy.}
\end{figure*}

Summarising, we have shown that the angular momentum of spiral galaxies, not only in its orientation but also in its
magnitude, is affected by the presence of a companion, once the galaxies are within the virial radius of their neighbour. 
Once galaxies cross into their virial radii, a slight but measurable decrease in $\lambda$ begins
to operate. A corresponding increase in alignment also appears, but only in going to very close systems
where the out-of-equilibrium situation makes determining values of $\lambda$ unfeasible.

In a previous work (Cervantes-Sodi \& Hernandez 2009) using a reduced sample of very well studied spiral galaxies,
two of us showed how the star formation rate has a marked dependence on $\lambda$,
it decreases when the value of $\lambda$ increases. If such tendency remains in perturbed systems, it could explain
the enhanced star formation present in interacting galaxies, as due to the decrease of the spin. Some studies (Lambas,
Tissera, Alonso \& Coldwell 2003; Ellison, Patton, Simard \& McConnachie 2008) based on observational samples coming
from large galaxy surveys, even show a gradual increase of the star
formation as the distance between the interacting galaxies decreases, result completely compatible
with our finding of the decrease of the value of $\lambda$ in such circumstances. More recently, Park \& Choi (2009),
studying the effects of galaxy interactions on galaxy properties, reported the enhancement of the star 
formation activity for target galaxies within the virial radius of the neighbour galaxy, when this
neighbour is a late type galaxy.

We explore the effects of the galaxy type of the nearest neighbour by splitting sample C in two according 
to the neighbours morphology, and plotting the value of $\lambda$ as a function of the normalised distance
between neighbouring galaxies $d/R_{vir}$, in Fig. 4 left panels. The top panel shows the behaviour when the nearest 
neighbour is a late type galaxy (2073 galaxies) and bottom panel when it is an early type one (1551 galaxies). 
The complementary
plots showing the difference in the position angle splitting sample D in two according
to the neighbours morphology, are shown in the right panels of Fig. 4; top panel for nearest
neighbours being late type (1325 galaxies) and bottom being early type (712 galaxies). If we look at the case
when the neighbour is a late type galaxy, we see that
within the errors, no effects are apparent when the normalised distance is larger than 1, with a slight decrease
in the values of $\lambda$ as the
normalized distance drops below $1R_{vir}$ and with an increase in the alignment or anti-alignment between
their position angles. The case when the neighbour galaxy is an early type one is more erratic
and the trend less evident, in both the magnitude of $\lambda$ and the spin alignment. It is
interesting to note that for the first bin, where the interaction is ongoing, 
we see much more significant disturbances in the value of $\lambda$, when both galaxies are spiral, probably due to the strong 
hydrodynamical effects of the interaction of two gaseous disks, which are absent when one of the two 
systems is of early type.

As an extension to the work presented here, it would be desirable to probe particularly high density environments,
to test how the situation tends to the picture seen at cluster scales (see Park \& Hwang 2009 for environmental effects
on cluster galaxies). Within the hierarchical picture
of structure formation, evolutionary timescales decrease with increasing ambient densities, which in turn
also increase with time. In this way, one could hope to recover an evolutionary sequence for any given
process, from a comparative study at zero redshift, extending over a large ambient density range. Also,
a fuller 3D alignment study might yield more extensive information. One must be cautious, e.g., on cluster 
scales, results of 2D alignment studies are sometimes modified when more extensive 3D analysis are
performed.

\section{Conclusions}

Using pairs of spiral galaxies of comparable size, we searched for correlations between their spins.
Using the spin parameter $\lambda$ to account for the magnitude of the angular momentum, and the 
position angle for its direction,
we found a weak but statistically relevant correlation between the spin magnitude
of neighbouring galaxies, but where unable to detect any tendency for alignment.

If we adopt the simplest version of the tidal torque theory to
explain the acquisition of angular momentum, we would expect a strong correlation for the spins
of spiral galaxies in a pair, if they had formed under the influence of the same tidal field. Our results imply the dominant 
presence of two complementary effects, the late formation of the galaxy pair, after the epoch of $\lambda$
acquisition was completed, and the blurring
of initial conditions due to the clumpy and irregular mass accretion of relatively extended minor merging events.

In going to the trends exhibited by the sample as a function of galaxy separation, we can clearly
see the effects of galaxy-galaxy interaction. These appear as soon as the galaxies cross into their 
virial radii, leading to a gradual decrease in the values of $\lambda$.

Our results do not diminishes the importance of the torquing at early stages of galaxy formation,
 but give us an insight into the important role
of later interactions in the overall evolution of the total angular momentum of the galaxies, as shown
by the results obtained from interacting systems, where the spin is visibly perturbed by the presence of a
nearby companion.

\section*{Acknowledgments}

The authors acknowledge the thorough reading of the original manuscript by the referee,
Binil Aryal, as helpful in reaching a clearer and more complete final version.

The work of B. Cervantes-Sodi was partially supported by a CONACYT scholarship.
The work of X. Hernandez was partially supported by DGAPA-UNAM grant no IN114107. 

CP acknowledges the support of the Korea Science and Engineering
Foundation (KOSEF) through the Astrophysical Research Center for the
Structure and Evolution of the Cosmos (ARCSEC).

Funding for the SDSS and SDSS-II has been provided by the Alfred P. Sloan
Foundation, the Participating Institutions, the National Science
Foundation, the U.S. Department of Energy, the National Aeronautics and
Space Administration, the Japanese Monbukagakusho, the Max Planck
Society, and the Higher Education Funding Council for England.
The SDSS Web Site is http://www.sdss.org/.

The SDSS is managed by the Astrophysical Research Consortium for the
Participating Institutions. The Participating Institutions are the
American Museum of Natural History, Astrophysical Institute Potsdam,
University of Basel, Cambridge University, Case Western Reserve University,
University of Chicago, Drexel University, Fermilab, the Institute for
Advanced Study, the Japan Participation Group, Johns Hopkins University,
the Joint Institute for Nuclear Astrophysics, the Kavli Institute for
Particle Astrophysics and Cosmology, the Korean Scientist Group, the
Chinese Academy of Sciences (LAMOST), Los Alamos National Laboratory,
the Max-Planck-Institute for Astronomy (MPIA), the Max-Planck-Institute
for Astrophysics (MPA), New Mexico State University, Ohio State University,
University of Pittsburgh, University of Portsmouth, Princeton University,
the United States Naval Observatory, and the University of Washington.


\begin{thebibliography}{99}

\bibitem{}Aryal B., Paudel S., Saurer W., 2007, MNRAS, 379, 1011

\bibitem{}Aryal B., Saurer W., 2006, MNRAS, 366, 438

\bibitem{}Avila-Reese V., Colin P., Gottl\"ober S., Firmani C., Maulbetsch C., 2005, ApJ, 634,51

\bibitem{}Bailin J., Steinmetz M., 2005, ApJ, 627, 647

\bibitem{}Barnes J., Efstathiou G.,1987, ApJ, 319, 575

\bibitem{}Berta Z. K., Jimenez R., Heavens A. F., Panter B. 2008, MNRAS, 391, 197

\bibitem{}Bett P., Eke V., Frenk C. S., Jenkins A., Helly J., Navarro J., 2007, MNRAS, 376, 215

\bibitem{}Boissier, S., Boselli, A., Prantzos, N., Gavazzi, G., 2001, MNRAS, 321, 733

\bibitem{}Bullock J. S., Dekel A., Kolatt T. S., Kravtsov A. V., Klypin A. A., Porciani C., 
Primack J. R., 2001, ApJ, 555, 240

\bibitem{}Cervantes-Sodi B., Hernandez X., Park C., Kim J., 2008, MNRAS, 863, 872

\bibitem{}Cervantes-Sodi B., Hernandez X., 2009, RevMexAA, 45, 75

\bibitem{}Cho J., Park C., 2009, ApJ, 693, 1045

\bibitem{}Choi Y., Park C., Vogeley M. S., 2007, ApJ, 658, 884

\bibitem{}Cole S., Lacey C., 1996, MNRAS, 281, 716

\bibitem{}Coutts A., 1996, MNRAS, 278, 87

\bibitem{}Croft R. A. C., Di Matteo T., Springel V., Hernquist L., 2009, MNRAS, 400, 43

\bibitem{}Davis A., Natarajan P., 2009, MNRAS, 393, 1498

\bibitem{}D'Onghia E., Navarro J. F., 2007, MNRAS, 380, L58

\bibitem{}Doroshkevich A. G., 1970, ApJ, 6, 320

\bibitem{}Ellison S. L., Patton D. R., Simard L., McConnachie A. W., 2008, ApJ, 135, 1877

\bibitem{}Fall S. M., Efstathiou G., 1980, MNRAS, 193, 189

\bibitem{}Faltenbacher A., Allgood B., Gottl\"ober S., Yepes G., Hoffman Y., 2005, MNRAS,
362, 1099

\bibitem{}Falthenbacher A., Li C., Mao S., van den Bosch F. C., Yang X., Jing Y. P.,
Pasquali, A., Mo H. J., 2007, ApJ, 662, L71

\bibitem{}Gardner J. P., 2001, ApJ, 557, 616

\bibitem{}Gott J. R., Rees M. J. 1975, A\&A, 45, 365

\bibitem{}Gott J. R., Thuan T. X., 1978, ApJ, 223, 426

\bibitem{}Gurovich S., McGaugh S. S., Freeman K. C., Jerjen H., Staveley-Smith L., 
De Block W. J. G., 2004, PASA, 21, 412

\bibitem{}Helou G., 1984, ApJ, 284, 471

\bibitem{}Hernandez X., Lee W. H., 2004, MNRAS, 347, 1304

\bibitem{}Hernandez X., Cervantes-Sodi B., 2006, MNRAS, 368, 351 

\bibitem{}Hernandez X., Park C., Cervantes-Sodi B., Choi Y., 2007, MNRAS, 375, 163 

\bibitem{}Hernquist L., 1993, ApJ, 409, 548

\bibitem{}Hoyle F., 1949, MNRAS, 109, 365

\bibitem{}Jimenez R., Padoan P., Matteucci F., Heavens A. F., 1998, MNRAS, 299, 123

\bibitem{}Kitzbichler M.G., Saurer W., 2003, ApJ, 590, L9

\bibitem{}Lambas D. G., Tissera P.B., Alonso M. S., Coldwell G., 2003, MNRAS, 346, 1189

\bibitem{}Maccio A. V., Dutton A. A., van den Bosch F. C., Morre B., Potter D., Stadel J., 2007, 
MNRAS, 378, 55

\bibitem{}Mo H. J., Mao S., White S. D. M., 1998, MNRAS, 295, 319

\bibitem{}Navarro J., Abadi M. G., Steinmetz M., 2004, ApJ, 613, L41

\bibitem{}Oosterloo T., 1993, A\&A, 272, 389

\bibitem{}Park C., Choi Y., 2005, ApJ, 635, L29

\bibitem{}Park C., Choi Y., 2009, ApJ, 691, 1828

\bibitem{}Park C., Gott J. R., Choi, Y.-Y., 2008, ApJ, 674, 784

\bibitem{}Park C., Hwang H. S., 2009, ApJ, 699, 1595

\bibitem{}Peebles P. J. E., 1969, ApJ, 155, 393

\bibitem{}Peirani S., Mohayaee R., de Freitas Pacheco J. A., 2004, MNRAS, 348, 921

\bibitem{}Pesta\~na, J. L. G., Cabrera J., 2004, MNRAS, 353, 1197

\bibitem{}Piontek F., Steinmetz M., 2009, MNRAS submitted (astro-ph/0909.4167)

\bibitem{}Pizagno J., et al. 2007, AJ, 134, 945

\bibitem{}Plionis M., Basilakos S., 2002, MNRAS, 327, L47

\bibitem{}Plionis M., Benoist C., Maurogordato S., Ferrari C., Basilako, S., 2003,
ApJ, 594, 144

\bibitem{}Porciani C., Dekel A., Hoffman Y., 2002, MNRAS, 332, 325

\bibitem{}Prantzos, N., Boissier S., 2000, MNRAS, 313, 338

\bibitem{}Schaefer, B. M., 2009, Int. J. Modern Phys. D, 18, 173

\bibitem{}Sharp N. A., Lin D. N. C., White S. D. M., 1979, MNRAS, 187, 287

\bibitem{}Shaw L. D., Weller J., Ostriker J. P., Bode, P., 2006, ApJ, 646, 815

\bibitem{}Slosar A., et al. 2009, MNRAS, 392, 1225

\bibitem{}Syer D., Mao S., Mo H. J., 1999, MNRAS, 305, 357

\bibitem{}Tonini C., Lapi A., Shankar F., Salucci P., 2006, ApJ, 638, L13

\bibitem{}Torlina L., De Propris R., West M. J., 2007, ApJ, 660, L97

\bibitem{}Unterborn C., Ryden B. S., 2008, ApJ, 687, 976

\bibitem{}van den Bosch F. C., 1998, ApJ, 507, 601

\bibitem{}Vitvitska M., Klypin A. A., Kravtsov A. V., Wechsler R. H., Primack J. R., Bullock J. S., 
2002, ApJ, 581, 799

\bibitem{}Wang Y, Park C., Yang X., Choi Y., Chen X., 2009, ApJ, 703, 951 

\bibitem{}Warren M. S., Quinn P. J., Salomon J. K., Zurek W. H., 1992, ApJ, 399, 405

\bibitem{}White S. D. M., 1984, ApJ, 286, 38

\bibitem{}Yang X., van den Bosch F. C., Mo H. J., Mao S., Kang X., Weinmann S. M., Guo Y., Jing Y. P., 2006
MNRAS, 369, 1293

\bibitem{}Zavala J., Okamoto T., Frenk C. S., 2008, MNRAS, 387, 364

\end{thebibliography}
\end{document}